\documentclass[
  journal=pasa,
  manuscript=research-paper, 
  year=2022,
  volume=40,
]{cup-journal}

\usepackage{microtype,siunitx,booktabs,amssymb,aas-macros}
\title{The origin of the spectral versus dynamical age discrepancy in radio galaxies}

\author{L. Jerrim}
\affiliation{School of Natural Sciences, Private Bag 37, University of Tasmania, Hobart, 7001 TAS, Australia}
\email[L. Jerrim]{larissa.jerrim@utas.edu.au}

\author{S. Shabala}
\affiliation{School of Natural Sciences, Private Bag 37, University of Tasmania, Hobart, 7001 TAS, Australia}
\alsoaffiliation{ARC Centre of Excellence for All Sky Astrophysics in 3 Dimensions (ASTRO 3D), Australia}

\author{R. J. Turner}
\affiliation{School of Natural Sciences, Private Bag 37, University of Tasmania, Hobart, 7001 TAS, Australia}

\author{P. Yates-Jones}
\affiliation{School of Natural Sciences, Private Bag 37, University of Tasmania, Hobart, 7001 TAS, Australia}
\alsoaffiliation{ARC Centre of Excellence for All Sky Astrophysics in 3 Dimensions (ASTRO 3D), Australia}

\author{M. Krause}
\affiliation{Centre for Astrophysics Research, University of Hertfordshire, College Lane, Hatfield, Herts AL10 9AB, UK}

\author{G. Stewart}
\affiliation{School of Natural Sciences, Private Bag 37, University of Tasmania, Hobart, 7001 TAS, Australia}

\author{C. Power}
\affiliation{International Centre for Radio Astronomy Research, University of Western Australia, 35 Stirling Highway, Crawley, WA 6009, Australia}
\alsoaffiliation{ARC Centre of Excellence for All Sky Astrophysics in 3 Dimensions (ASTRO 3D), Australia}

\doi{10.1017/pasa.2020.32}

\received {dd Mmm YYYY}
\revised  {dd Mmm YYYY}
\accepted {dd Mmm YYYY}
\published{dd Mmm YYYY}


\begin{document}

\begin{abstract}

We investigate the effect of turbulent magnetic fields on the observed spectral properties of synchrotron radio emission in large-scale radio galaxy lobes. We use three-dimensional relativistic magnetohydrodynamic simulations of fast, high-powered jets to study the structure of the lobe magnetic fields and how this structure affects the radio spectrum of the lobes. It has previously been argued that lobe ages inferred from radio spectra underestimate the true ages of radio galaxies due to re-acceleration of electrons in the lobe, mixing of electron populations, or the presence of turbulent magnetic fields in the lobes. We find that the spectral ages with and without accounting for the lobe magnetic field structure are consistent with each other, suggesting that mixing of radiating populations of different ages is the primary cause of the underestimation of radio lobe ages. By accounting for the structure of lobe magnetic fields, we find greater spectral steepening in the equatorial regions of the lobe. We demonstrate that the assumptions of the continuous injection, Jaffe-Perola, and Tribble models for radio lobe spectra do not hold in our simulations, and we show that young particles with high magnetic field strengths are the dominant contributors to the overall radio lobe spectrum.

\end{abstract}

\section{Introduction}

Active Galactic Nuclei (AGN) jets can produce large-scale radio lobes that provide feedback to their galaxy group or cluster environment \citep{mcnamara_heating_2007}. Understanding the energetics associated with these objects is critical for building accurate models of galaxy formation and evolution over cosmic time \citep[e.g.][]{sijacki_unified_2007, raouf_feedback_2019, talbot_blandford-znajek_2021}. This requires understanding the total power output of these objects and the timescales over which they operate.

For large-scale Fanaroff-Riley (FR) class II sources \citep{fanaroff_morphology_1974}, there are two main methods of estimating the age of a radio lobe. The first is through using a dynamical model \citep[e.g.][]{kaiser_self-similar_1997}. The second is through using the radio spectrum of the lobe to estimate the age of the emitting particles; high energy particles have greater radiative losses, so older sources will have a steeper spectrum \citep{myers_synchrotron_1985}. These estimates rarely agree, and can be up to an order of magnitude different in some cases, where spectral age estimates are lower than dynamical age estimates \citep{eilek_how_1996, blundell_spectra_2000, harwood_spectral_2013}. For example, \citet{harwood_spectral_2015} find a maximum spectral age of $5.50^{+0.95}_{-0.4}$ Myr, and a rough dynamical age estimate of $9$ Myr for radio source 3C 438. This difference may be in part due to errors in the dynamical model or assumed magnetic field strength, especially for older, simpler models \citep[see e.g.][for a review]{turner_dynamics_2023}. 
\\

There are several different models that describe the shape of a lobe's radio spectrum, either in its entirety or for each individual beam of a resolved source. The main models used are the continuous injection \citep[CI;][]{kardashev_nonstationarity_1962, pacholczyk_radio_1970}, Jaffe-Perola \citep[JP;][]{jaffe_dynamical_1973}, and the Kardashev-Pacholczyk \citep[KP;][]{kardashev_nonstationarity_1962, pacholczyk_radio_1970} models. The CI, JP, and KP models each have a different set of assumptions associated with them, so they each have their strengths and weaknesses in modelling radio lobe emission. 

The CI model assumes a continuous injection of emitting particles over the source time, and is therefore most logically applied to the entirety of the emission for an active source. The CI model is not complete as it does not account for adiabatic losses. While the functional form of CI model spectra can be consistent with lobe spectra evolved under full (i.e. adiabatic, synchrotron and inverse Compton) losses dependent on a weakening magnetic field over time \citep{turner_raise_2018_II}, the parameters characterising such a spectrum will not have a straightforward physical interpretation.

By contrast, the JP and KP models both assume there is a single injection of emitting particles that produces a power law distribution of energies, and thus should only be applied to plasma shock-accelerated at approximately the same time. The JP and KP models differ in how the pitch angles of the emitting particles are treated; they are either assumed to scatter (JP) or to maintain the same pitch angle over time (KP). The KP model is generally not thought to be physically plausible as inhomogeneities in the magnetic field structure will scatter the pitch angle of the emitting particles \citep{tribble_radio_1993, turner_raise_2018_III}, and it is known that AGN jet-inflated lobes have complex magnetic field structures \citep[e.g.][]{mukherjee_simulating_2020, mullin_observed_2008}. 

Despite the complexity of radio lobe magnetic fields, the standard forms of the CI, JP, and KP models all assume both a spatially and temporally uniform magnetic field strength. These assumptions are in general violated, as radio lobes have turbulent magnetic fields that evolve over time due to jet dynamics \citep[e.g.][]{mukherjee_simulating_2020, meenakshi_polarization_2023, jerrim_faraday_2024}, which will change the synchrotron losses (and therefore the radio spectrum) over time. \citet{turner_raise_2018_II} have previously demonstrated that the vast majority of emission observed at common radio frequencies is produced by the youngest particles in the spectrum, which will have minimal evolution in the magnetic field over time, so the assumption of a temporally uniform lobe magnetic field is likely to be only weakly violated.

The \citet{tribble_radio_1991} model accounts for localised spatial variation in the lobe magnetic field by assuming that the magnetic field components follow a Gaussian random field with zero mean so that the magnetic field strength follows a Maxwell-Boltzmann distribution. Simulations of RMHD jets on scales of $\lesssim 10$ kpc show that this is generally a valid assumption \citep{mukherjee_simulating_2020}, but it is unclear whether this result holds in larger cocoons with more turbulence. The CI, JP, and KP models can all be modified to assume this magnetic field distribution \citep[e.g.][]{harwood_spectral_2013}, and are known as the TCI, TJP, and TKP models respectively.

The shape of the radio lobe spectrum (either integrated or spatially resolved) is related to its spectral age through the so-called `spectral break' in the spectrum. A spectrum `ages' over time due to synchrotron and inverse-Compton losses, resulting in a steeper spectrum and higher spectral index value ($\alpha$, where $S_\nu \propto \nu^{-\alpha}$). The spectral break frequency $\nu_{b}$ is a characterisation of where the spectrum steepens, which occurs over a range of frequencies \citep{blundell_spectra_2000}. The location of the spectral break is time-dependent since high energy particles lose energy more rapidly than lower energy particles. The spectral age ($\tau$) can be related to the magnetic field strength $B$ and $\nu_{b}$ as follows \citep{hughes_beams_1991,turner_raise_2018_III}:

\begin{equation}
\label{eqn:spectral-age}
    \tau = \frac{v B^{1/2}}{B^2 + B_{\rm ic}^2} \left[\nu_b (1 + z) \right]^{-1/2},
\end{equation}

where $z$ is the redshift of the source, $B_{\rm ic} = 0.318 (1 + z)^2$ nT is the magnitude of the magnetic field equivalent to the cosmic microwave background energy density, and the constant $v$ for the JP model is given by:

\begin{equation}
    v = \left( \frac{243 \pi m_e^2 c^2}{4 \mu_0^2 e^7} \right)^{1/2},
\end{equation}

where the symbols have their usual meaning. The \textsc{synchrofit} Python library \citep{quici_selecting_2022} uses this relationship to parameterise the radio spectrum model in terms of $\nu_{b}$, simplifying the spectral age calculation by removing $B$ as a variable. In contrast, the \textsc{BRATS} method \citep{harwood_spectral_2013,harwood_spectral_2015} parameterises the radio spectrum in terms of $B$, $B_{\rm ic}$, and $\tau$. The relationship shown in Equation \ref{eqn:spectral-age} is not valid when a broken power-law method is used to fit the break frequency in a non-self-consistent manner, rather than a complete CI, JP, or KP model, which will feature continuous spectral curvature and yield $\nu_b$ as a characteristic parameter of the model (see \ref{section:appendix} for more detail).

Any processes that affect the synchrotron spectrum will affect the spectral age estimate. \citet{hardcastle_synchrotron_2013} investigated the role of magnetic field structure on radio spectra, and found that the spectral break moves to higher frequencies at low redshift ($z = 0$), thus increasing the spectral age estimate; this analytic result does not account for the dynamics of the jet and its cocoon. Additionally, \textit{in situ} acceleration and/or mixing of emitting particle populations with different ages will impact spectral age estimates. \citet{turner_raise_2018_II} have shown that particle mixing in FR-II type radio lobes can reduce the spectral age while maintaining a gradient of young to old spectral ages in the backflow from the hotspots using an analytic model informed by hydrodynamic simulations. 

The specific differences between the spectral properties of radio lobes can be studied using numerical simulations. \citet{hardcastle_numerical_2014} and \citet[][hereafter referred to as Paper I]{jerrim_braise_2025} discuss how the complex magnetic field structure in the jet cocoon affects the polarised radio emission in the lobes. However, neither study discusses in detail how the turbulent magnetic field affects the radio spectrum. In this paper, we study three-dimensional relativistic magnetohydrodynamic simulations of FR-II-like sources using the PRAiSE \citep{yates-jones_praise_2022} and BRAiSE (Paper I) synthetic radio emission methods to investigate the influence of turbulent magnetic fields on radio lobe emission and how this impacts spectral age measurements. 

The paper is structured as follows. We introduce the simulations discussed in this paper (also used in Paper I) in Section \ref{section:simulations_specage}. We discuss the dynamics of our magnetised jets in Section \ref{section:dynamics_specage} with a focus on their stability. In Section \ref{section:spectra}, we compare the spectral properties of three different spectral models for our highly magnetised jet simulations. The influence of the different emission methods on the resulting spectral age estimates is discussed in Section \ref{section:dyn-spec-ages}. We discuss the roles of the mixing of particle populations of different ages and the magnetic field in radio lobe spectra in Section \ref{section:discussion_specage} and finish by summarising our findings in Section \ref{section:conclusions_specage}.

\section{Simulations}
\label{section:simulations_specage}

In Paper I, we presented four simulations of relativistic jets: one relativistic hydrodynamic (RHD) simulation, and two relativistic magnetohydrodynamic (RMHD) simulations with different jet magnetic field strengths in a group environment, and one RMHD simulation in a cluster environment. We focused on the dynamics in the RMHD simulations and how this informs the differences in the polarisation properties of their simulated sources. In this paper, we further explore the differences between calculating the synchrotron emissivity from the magnetic field (BRAiSE; Paper I) to previous methods using the pressure \citep[PRAiSE;][]{turner_raise_2018_II, yates-jones_praise_2022}. For consistency, we keep the nomenclature of the simulations the same as in Paper I; the simulations discussed in this paper and their relevant parameters are listed in Table \ref{tab:sim_list_specage}. We refer the reader to Paper I for the full details of these simulations and summarise briefly below.

Our simulations were carried out using the \textsc{PLUTO} astrophysical fluid dynamics code \citep[version 4.3;][]{mignone_pluto_2007} using the RMHD physics and RHD physics modules. For the environment, we used radially-averaged galaxy group and cluster environments from \textsc{The Three Hundred} project \citep{cui_three_2018}. These environments are then combined with a generated magnetic field that has a minimum wavenumber $k_{\rm min} = 0.015$ kpc$^{-1}$ and average magnetic field strengths of $0.1 \mu$G and $1 \mu$G in the group and cluster environments respectively, following the method presented in \citet{jerrim_faraday_2024}.

To simulate FR-II like sources, we injected fast, relativistic jets with total one-sided jet power $Q_j = 1 \times 10^{38}$ W, jet Lorentz factor $\Gamma = 5$, ratio of kinetic to thermal energy in the jet $\chi = 100$, and jet half-opening angle $\theta = 15\deg$. The jet magnetic field is toroidal \citep{jerrim_faraday_2024}, with an injection strength as detailed in Table \ref{tab:sim_list_specage}. The density and pressure of each jet are adjusted so that the total energy flux from the jet is held constant for all three simulations (Paper I). We primarily study the two simulations with the higher jet magnetic field strength in this paper since simulation RAG-B16 was shown in Paper I to be an FR-I-like source; however, we do refer to simulation RAG-B16 as a point of comparison in Section \ref{section:dynamics_specage}.

\begin{table}
	\centering
	\caption[Simulation parameters]{Parameters of the simulations discussed in this paper. $B$ is the initial magnetic field strength in the `cap' of the injection cone. $\rho$ and $p$ are the density and pressure in the jet, respectively. $Q_{B}/Q_{k}$ is the ratio of magnetic to kinetic energy flux in the jet. The letters in the `Env' column correspond to the type of environment the simulation has; `G' for group and `C' for cluster.}
	\label{tab:sim_list_specage}
    \begin{tabular}{llllll}
    \hline
         Name & $B$ ${\rm (} \mu {\rm G)}$ & $\rho$ ${\rm (g/cm^3)}$ & $p \; {\rm (Pa)}$  & $Q_{B}/Q_{k}$ & Env \\
    \hline
         RAG-B16 & $16$ & $3.94 \times 10^{-30}$ & $1.44 \times 10^{-11}$ & 0.0003 & G \\
         RAG-B327 & $327$ & $3.57 \times 10^{-30}$ & $1.30 \times 10^{-11}$ & 0.13 & G \\
         RAC-B327 & $327$ & $3.57 \times 10^{-30}$ & $1.30 \times 10^{-11}$ & 0.13 & C \\
    \hline
    \end{tabular}
\end{table}

\section{Dynamics}
\label{section:dynamics_specage}

In Paper I, we focused on the dynamical features that affect the synthetic radio emission. The shock structures shown in Paper I indicate that both the RAG-B327 and RAC-B327 simulations have FR-II-like morphology, as the freshly shocked particles form a hotspot at the end of each jet. We found an additional feature in both simulations that appears similar to a flaring point; this 'brightening point' is caused by Kelvin-Helmholtz instabilities along the jet channel. We briefly discussed the kink instabilities present in both jets, which cause the forward flow to deviate from the jet axis and affect jet propagation. These instabilities disperse the magnetic energy in the hotspot region. We discuss the stability of the jets in more detail here.

\begin{figure}
    \centering
    \includegraphics[width=\textwidth]{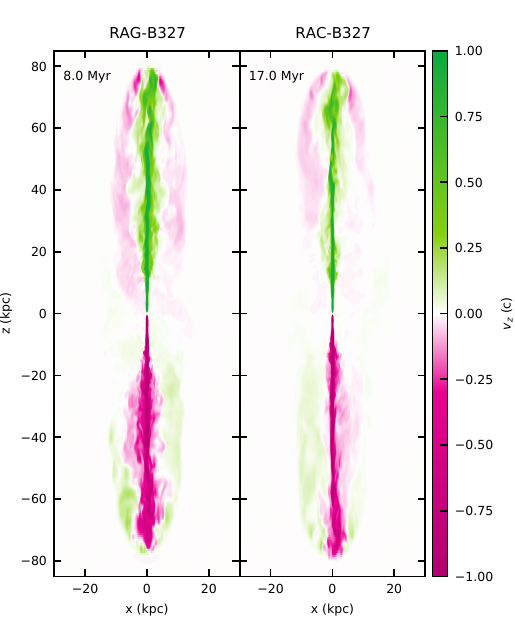}
    \caption[Midplane slices at $y = 0$ kpc of the $z-$component of the velocity]{Midplane slices of the $z$-velocity at $y = 0$ for each simulation, with increasing environment density from left to right. Simulation RAG-B327 is shown at $8$ Myr and simulation RAC-B327 is shown at $17$ Myr. The backflow is shown in pink for the upper jets and green for the lower jets.}
    \label{fig:vel-slices}
\end{figure}

In Fig. \ref{fig:vel-slices} we plot the midplane slice $z$-velocity for our high jet magnetic field strength simulations at the same total source length (roughly $160$ kpc). This velocity component is parallel to the jet axis and shows the strength of the backflow. This backflow occurs as a result of the interaction between the jet and the swept-up ambient medium; a more stable jet will drive a stronger backflow that fills out the jet cocoon. The backflow is weaker in simulation RAC-B327, due to the differences in the pressure contrast between the hotspot and the lobe. In this simulation, the higher density environment slows down jet propagation, and therefore the jet has injected more energy into the lobe and increased the lobe pressure relative to the group simulations (shown at earlier times). Both jets in these simulations are affected by kink instabilities that cause the jet to bend rather than propagate directly down the $z$-axis. This effect can clearly be seen near the jet head, where the jet is beginning to be disrupted by the instabilities.

\begin{figure}
    \centering
    \includegraphics[width=\textwidth]{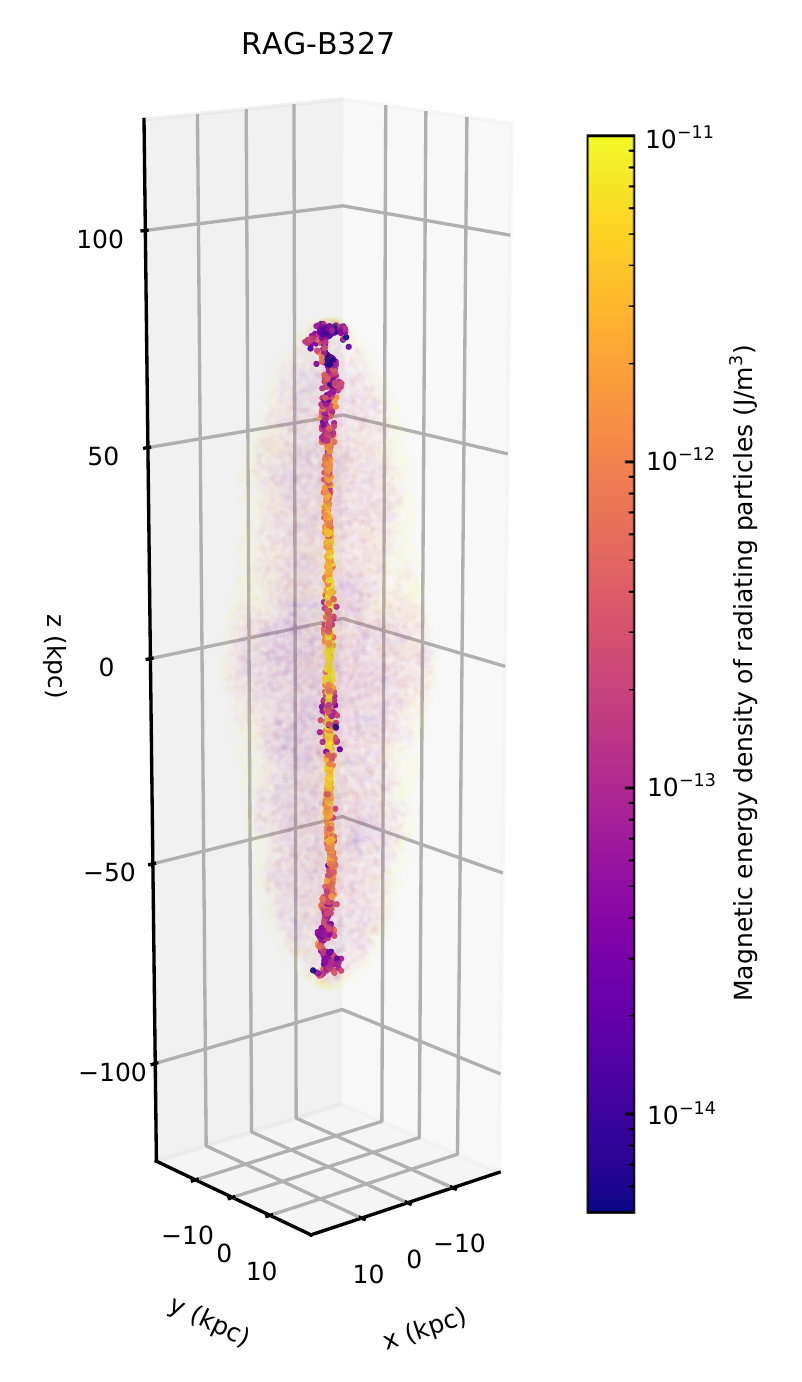}
    \caption[Three-dimensional particle scatterplot of the magnetic energy density in the radiating particles]{3D illustration of the particle magnetic energy density in simulation RAG-B327 at $8$ Myr, demonstrating the dispersion of magnetic energy in the jet head. Particles with a fluid tracer value $> 0.2$ are plotted with full opacity; particles with lower fluid tracer values are plotted with low opacity to show the lobe shape.}
    \label{fig:3d-ub-RAG}
\end{figure}

As mentioned in Paper I, this movement of the jet impacts the magnetic energy in the jet head region. In Fig. \ref{fig:3d-ub-RAG}, we plot the magnetic energy density in the radiating particles for simulation RAG-B327. The upper jet in simulation RAG-B327 has a kink instability at $z \simeq 50$ kpc, after which the magnetic energy density drops by two orders of magnitude as it reaches the end of the jet. The bending of the jet is clearly seen as this instability disrupts the jet flow and disperses the magnetic energy.

\begin{figure*}
    \centering
    \includegraphics[width=\textwidth]{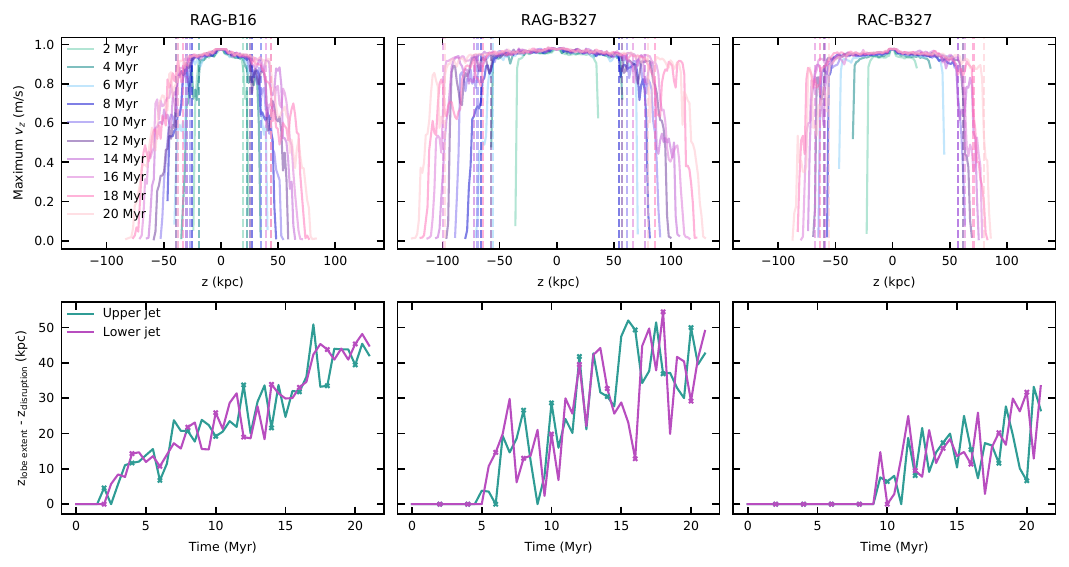}
    \caption[Jet disruption: maximum particle velocity along the jet axis and distance between the disruption point and lobe edge over time]{Top row: maximum absolute z-velocities of particles within $\pm 1$ kpc of the z-axis along the z-axis for each simulation. Bottom row: distance between the disruption point and the end of the lobe over time. Simulations RAG-B16, RAG-B327, and RAC-B327 are shown from left to right. The dashed vertical lines in the top row correspond to the disruption point at each time shown. The crosses in the bottom row correspond to the times shown in the top row. }
    \label{fig:disruption}
\end{figure*}

Jet disruption can be quantified using a velocity gradient to find where the jet slows down. In the top row of Fig. \ref{fig:disruption}, we plot the maximum absolute velocity along the $z$-axis of particles within $\pm 1$ kpc of the $z$-axis. Along each jet, the first location with a significant velocity gradient corresponds to the disruption point, i.e. the first location where the jet slows down significantly from its initial velocity. The distance between this point and the end of the jet lobe is displayed in the bottom row. Simulation RAG-B16 is included to demonstrate the disruption in an FR-I-like source as compared to the FR-II-like sources in simulations RAG-B327 and RAC-B327. 

At early times in both high jet magnetic field simulations, the jet is stable. In simulation RAG-B327, both jets are disrupted by $5.5$ Myr; however, in simulation RAC-B327, both jets are disrupted at $9.5$ Myr. Once the jets are disrupted, the distance between the jet origin and the disruption point does not change appreciably over time for all three simulations. In general, this disruption point does not spatially correlate to the brightening point, as seen in the shock analysis in Fig. 3 of Paper I, since this point is primarily caused by Kelvin-Helmholtz instabilities along the jet, rather than the kink instabilities that bend the jet.

Previous studies have shown that fast, strongly magnetised jets are more stable \citep{mukherjee_simulating_2020, rossi_different_2024}. We confirm this result with simulations RAG-B16 and RAG-B327; when the magnetisation of the jet is increased, for the same total jet power, jet collimation is retained for a greater distance along the jet. Our simulations do not include all the physical processes that can act to stabilise an RMHD jet present in our simulations \citep[e.g. jet rotation;][]{bodo_linear_2016}, however, the jets in simulations RAG-B327 and RAC-B327 inflate the cocoons with backflow from the hotspot, producing FR-II-like radio lobes.

\vfill\null

\begin{figure}
    \centering
    \includegraphics{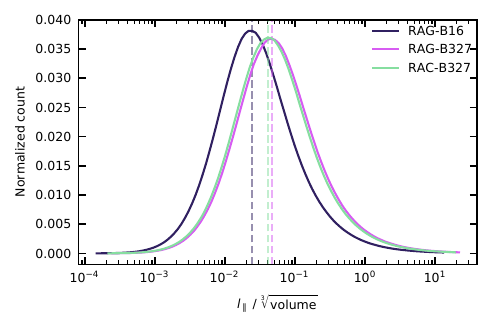}
    \caption[Distributions of the field curving length scale]{Distributions of the field curving length scale for simulations RAG-B16, RAG-B327, and RAC-B327, plotted at times corresponding to a total source length of roughly $160$ kpc ($20$ Myr/$8$ Myr/$17$ Myr respectively). This length scale has been normalised by the cube root of the volume to account for differences in morphology. The jet has been removed from this calculation using a tracer threshold $< 0.1$, and the lobe is defined using a tracer threshold of $> 10^{-4}$.}
    \label{fig:b-l-para-histograms}
\end{figure}

The disruption of the jet can also be characterised by the turbulent magnetic field structures in the lobes. In Fig. \ref{fig:b-l-para-histograms}, we plot the field curving length scale normalised by the cube root of the source volume to study the structure of the magnetic fields in the lobes. This field curving length scale is the inverse of Schekochihin's magnetic wave number and a measure for the length scale over which magnetic fields are significantly bent. It is defined as \citep{schekochihin_simulations_2004, bodo_symmetries_2011}:

\begin{equation}
    l_\parallel = \left[ \frac{|\mathbf{B}|^4}{|(\mathbf{B}\cdot\nabla)\mathbf{B}|^2} \right]^{1/2}.
\end{equation}

The distribution of this field curving length scale in the lobes depends on the stability of the jet. \citet{mukherjee_simulating_2020} show for small-scale ($< 10$ kpc) jets that unstable jets tend to be dominated by smaller scale lengths in the lobes compared to stable jets. We confirm this result with our large-scale ($> 100$ kpc) jets, where the distribution of field curving length scales is skewed towards longer length scales in simulations RAG-B327 and RAC-B327 (FR-II-like cocoons) compared to simulation RAG-B16 (an FR-I-like cocoon). Despite simulations RAG-B327 and RAC-B327 exhibiting disrupted jets, this field curving length scale demonstrates that these simulations do not have true lobed FR-I-type morphology. This quantifies that lobe turbulence is fed on relatively larger scales in the B327 simulations with correspondingly, on average, larger eddies.  

\begin{figure}
    \centering
    \includegraphics{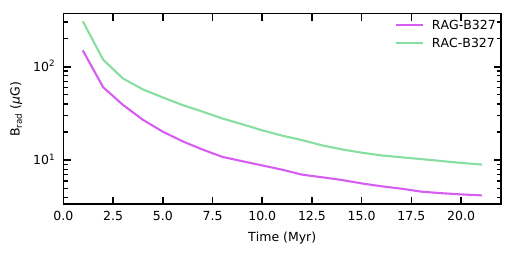}
    \caption[Average lobe magnetic field strength over time]{Average magnetic field strength in the lobes for simulations RAG-B327 and RAC-B327 over time. The average is a volume-weighted average of the logarithm of the magnetic field strength in the radiating particles in the lobe (i.e, $10^{\sum log(B_{\rm rad}) dV / \sum dV}$).}
    \label{fig:lobeB-avg}
\end{figure}

As in Paper I, we consider the sub-grid dilution of the lobe magnetic field strength due to the mixing of the lobe plasma with thermal particles from the ambient medium. That is, we expect these two fluids to remain somewhat segregated on scales smaller than those captured by our grid resolution, and thus the magnetic field strength should not be averaged across these two populations, as is done in the hydrodynamic simulations. We account for this sub-grid effect by scaling the simulation magnetic field strength by the volume filling factor of jet material to obtain the magnetic field strength of the radiating lobe plasma. The volume filling factor $f$ is:

\begin{equation}
\label{eqn:volumefill_specage}
    f = \frac{1}{1 + \frac{m_r}{m_{\rm th}} \left( \frac{1}{\rm trc} - 1 \right)},
\end{equation}

where $m_{r}$ is the mass of lobe plasma associated with each radiating particle (i.e., the sum of electron mass and positively charged particle mass), $m_{th}$ is the mass of the thermal particles, and ${\rm trc}$ is the passive fluid tracer value. For an electron-proton plasma as assumed in Paper I, $m_r / m_{th} = 1$ and the scaling simplifies to $B_{\rm rad} = B/\sqrt{f} = B/\sqrt{\rm trc}$. The magnetic field of the radiating particles in the lobe agrees with observations of FR-II radio lobes, which estimate magnetic field strengths of $1 - 20 \mu$G for sources with lobe sizes around $100-300$ kpc \citep[e.g.][]{ineson_representative_2017, turner_raise_2018_III}. For example, in simulation RAG-B327 at 8 Myr (160 kpc total length), the average magnetic field strength for the radiating particles is around $11 \mu$G. This is a volume-weighted average of the logarithm of the magnetic field strength to approximate the peak of the magnetic field distribution, which is approximately normally distributed in log space. We plot the average lobe magnetic field strength found in this manner for simulations RAG-B327 and RAC-B327 in Fig. \ref{fig:lobeB-avg}.

We note that there will be some reduction in the lobe magnetic energy due to numerical dissipation. Simulations of MHD turbulence on uniform grids have found that numerical dissipation has a greater effect for worse grid resolution \citep[e.g.][]{kritsuk_comparing_2011, shivakumar_numerical_2025}. However, the simulations presented here are on a stretched grid, so the changing resolution across the simulation grid will impact when and where the magnetic energy in the lobes is being dissipated. The field curving length scale in the lobes (as shown in Fig. \ref{fig:b-l-para-histograms}) range over about four orders of magnitude, so there will be a wide range of magnetic dissipation rates throughout the lobe. To avoid uncertainties due to numerical dissipation, \citet{mattia_resistive_2023} have conducted resisitive RMHD jet simulations. By including an explicit resistivity term, the authors are able to avoid excess dissipation by ensuring that numerical resistivity is not dominant \citep{thornber_implicit_2007}, however, the authors use a physics module for PLUTO \citep{mignone_fourth-order_2024} that has not yet been publicly released. We defer the investigation of resistive RMHD jet simulations to future studies.

We find that test simulations of the magnetised group and cluster environments used in this paper have shown that the half-life of magnetic energy on the stretched simulation grid is $13.0$ Myr and $15.2$ Myr respectively. We note that a small fraction of this magnetic energy loss is converted to kinetic and thermal energy through gas motions, but the remainder is lost due to numerical effects. By contrast, the adiabatic expansion of the lobe (in simulations with jets) reduces the lobe magnetic field strength much more rapidly than the rate of numerical dissipation, halving every 2 Myr or faster in both simulations (see Figure \ref{fig:lobeB-avg}). As the numerical dissipation occurs at a lower rate, the magnetic energy losses should not have a significant effect on our results. Further detailed investigation of numerical dissipation on stretched grids is beyond the scope of this study.

\section{Modelling FR-II lobe spectra}
\label{section:spectra}

We generate synthetic surface brightness images using two methods: with the magnetic field directly (BRAiSE; Paper I) and with the lobe pressure as a proxy for the magnetic field \citep[PRAiSE;][]{yates-jones_praise_2022}. The synthetic radio emission is calculated using passive Lagrangian tracer particles that are advected with the fluid on the simulation grid. We use an injection index of $\alpha_{\rm inj} = 0.6$ and minimum and maximum Lorentz factors $\gamma_{\rm min} = 500$ and $\gamma_{\rm max} = 10^5$, consistent with \citet{yates-jones_praise_2022}. We calculate the surface brightness images in the plane of the sky convolved with a 2D Gaussian beam with FWHM of $2.97$ arcsec at a redshift of $z = 0.05$. 

To study the interplay between particle mixing and turbulent magnetic fields on the spectral properties of FR-II-like radio lobes, we construct three models. In order of lowest to highest complexity, we have:

\begin{enumerate}
    \item Mixing only: calculates the radio emission using the median Lagrangian tracer particle pressure for all particles in the lobe. This model assumes that mixing of particle populations with different ages is the only contributor to how `young' a spectrum appears.
    \item Mixing + pressure (p) turbulence: calculates the radio emission using pressure \citep[i.e. PRAiSE;][]{yates-jones_praise_2022}. The inhomogeneities in the lobe pressure are a contributor to changes in the radio spectrum in addition to particle mixing.
    \item Mixing + magnetic (B) turbulence: calculates the radio emission using the full magnetic field information (i.e. BRAiSE; Paper I). This accounts for the contribution of the turbulent lobe magnetic field to changes in the radio spectrum as well as particle mixing.
\end{enumerate}

For the non-magnetic emission models, we use the equipartition factors as shown in Fig. 5 of Paper I to map pressure to the magnetic field energy density. We study the effect of each of our three emission models on the surface brightness, spectral energy distribution, and spectral index of the same simulated sources. We then quantify the effect of these models on the derived spectral age.

\subsection{Surface brightness}
\label{section:sb}

\begin{figure*}
    \centering
    \includegraphics{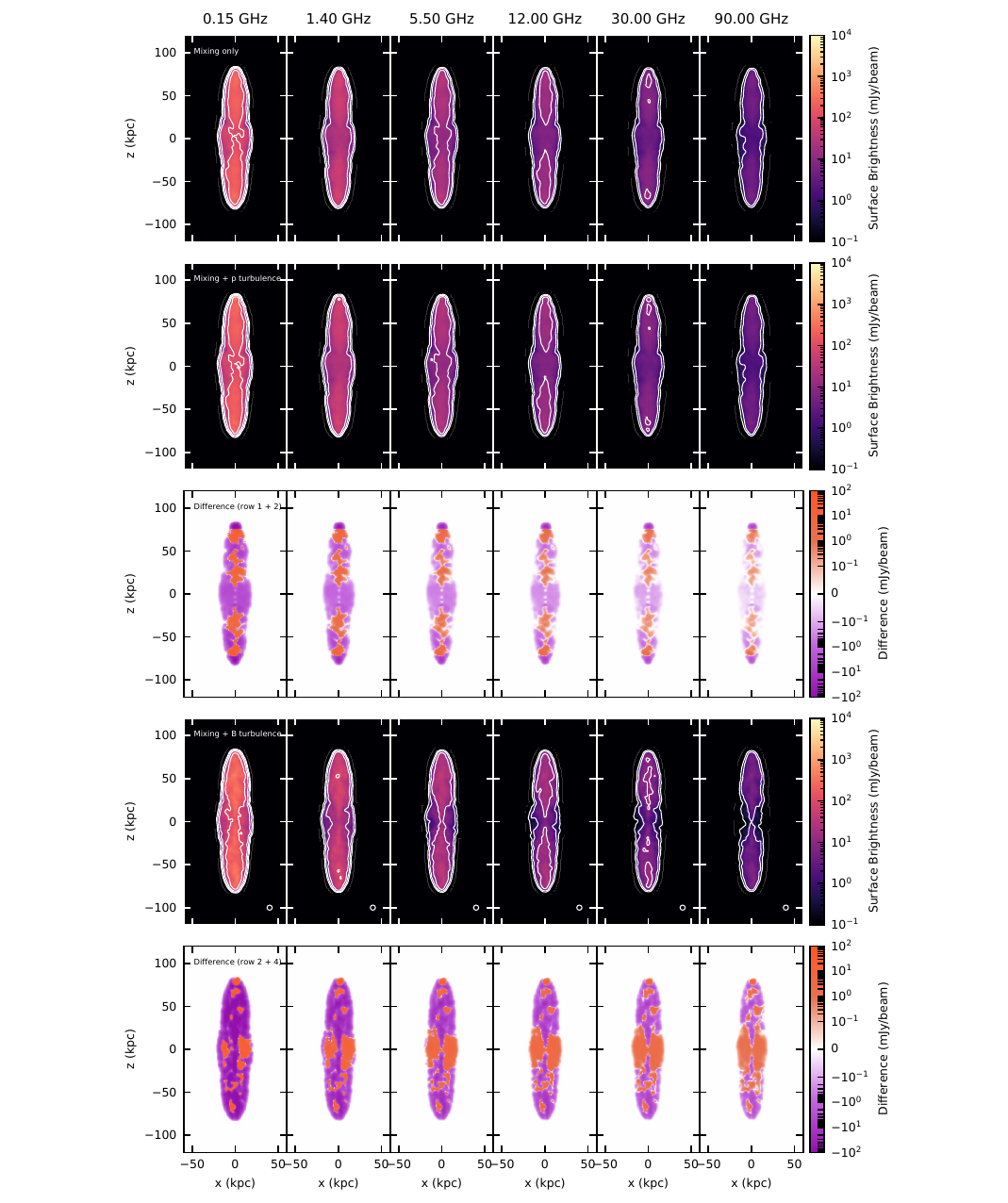}
    \caption[Surface brightnesses for the three spectral models for simulation RAG-B327]{Synthetic surface brightness images (Stokes' I) for each of the spectral models for simulation RAG-B327. Each model is shown at $0.15, 1.4, 5.5, 12.0, 30.0,$ and $90.0$ GHz. Contours are at $0.1, 1, 10, 100, 1000,$ and $10000$ mJy/beam. The beam FWHM is $2.97$ arcsec and the sources are simulated at a redshift of $z = 0.05$. The difference between the `mixing only' and `mixing + pressure turbulence' images is shown in the third row. The difference between the `mixing + pressure turbulence' and `mixing + magnetic turbulence' images is shown in the fifth row.}
    \label{fig:sb-RAG-B327}
\end{figure*}

\begin{figure*}
    \centering
    \includegraphics{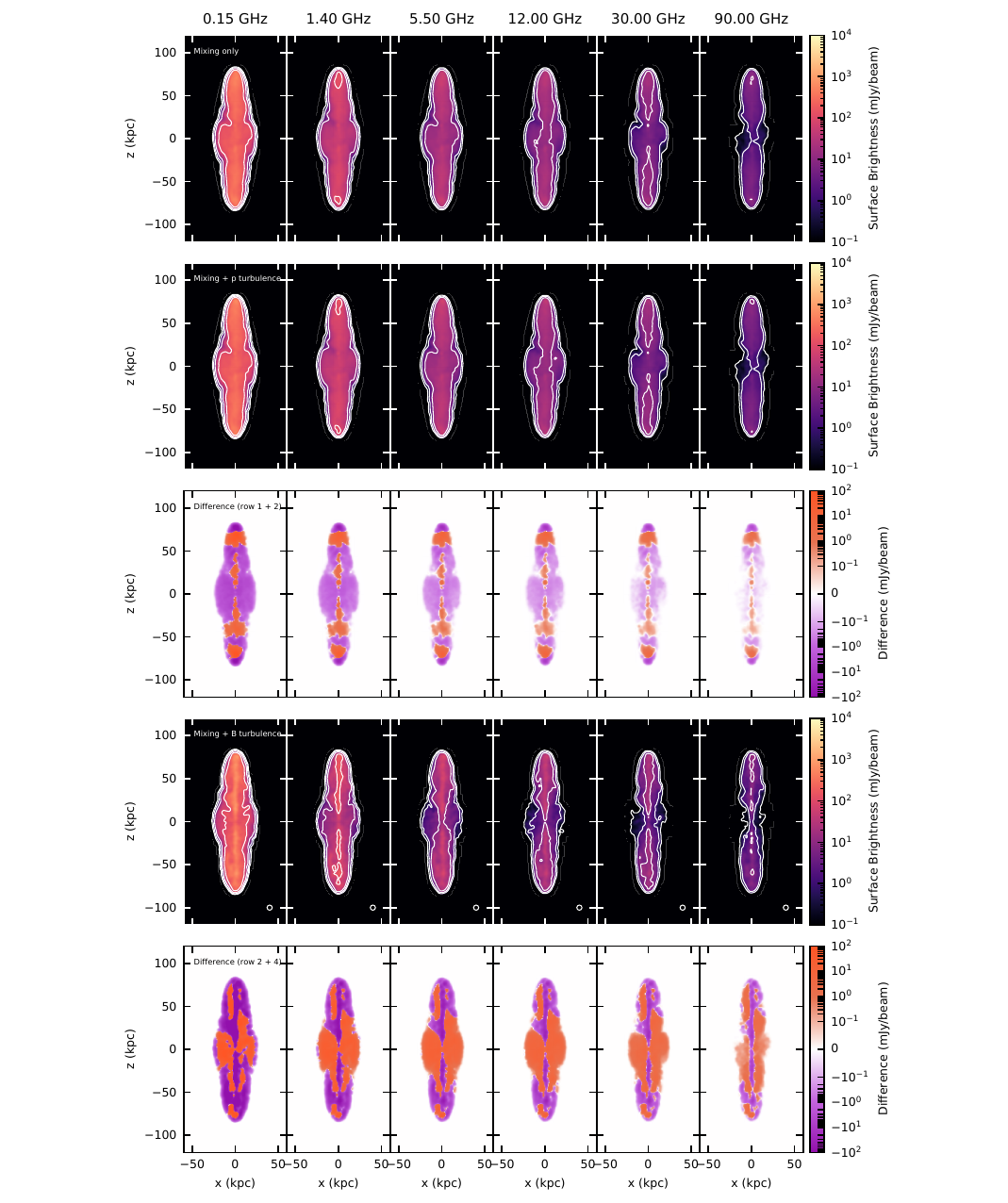}
    \caption[Surface brightnesses for the three spectral models for simulation RAC-B327]{Synthetic surface brightness images (Stokes' I) for each of the spectral models for simulation RAC-B327. Frequencies, contours and difference rows are as in Fig. \ref{fig:sb-RAG-B327}.}
    \label{fig:sb-RAC-B327}
\end{figure*}

We plot the surface brightness for each of the three models in Figs. \ref{fig:sb-RAG-B327} and \ref{fig:sb-RAC-B327} for simulations RAG-B327 and RAC-B327 respectively. For both of these simulations, there are minimal differences between the `mixing only' and `mixing + pressure turbulence' models. The differences between these models are similar in both the group and cluster simulations: the `mixing only' model is brighter in the brightening regions and just before the hotspots in both the upper and lower lobes. However, in the equatorial regions of the lobe and the hotspots, the `mixing + pressure turbulence' model is brighter. These differences are due to the change in pressure histories of the particles over time; for the `mixing only' model, the median pressure at each time is assumed for all of the radiating particles.

As in Paper I, the surface brightness images for the `mixing + magnetic turbulence' (BRAiSE) model show a different spatial distribution of radio emission. For both simulations, the jets are brighter, and the equatorial regions are dimmer for the `mixing + magnetic turbulence' model compared to the `mixing + pressure turbulence' model. At the highest frequency shown ($90$ GHz) the magnetic model displays greater losses about halfway along each lobe, resulting in more `pinched' lobes for both simulations.

As demonstrated in Paper I, the `mixing + magnetic turbulence' model (i.e. BRAiSE) produces bright emission along the jet and its associated brightening region (caused by Kelvin-Helmholtz instabilities). For simulations RAG-B327 and RAC-B327 at the times shown in Figs. \ref{fig:sb-RAG-B327} and \ref{fig:sb-RAC-B327}, the jet contributes approximately $16$ and $19$ percent of the emission at $0.15$ GHz in BRAiSE respectively (this is unaffected by particle resolution, as discussed in Paper I). This result is not in line with observations of FR-II sources, where the jet prominence (i.e. the amount of total source emission attributed to the jet) is $\lesssim 0.1$ percent \citep{mullin_observed_2008}. This bright emission associated with the jet and its brightening region are not seen for either the `mixing only' or 'mixing + pressure turbulence' models (using PRAiSE) since the pressure is not significantly higher in these regions compared to the rest of the lobe, whereas the magnetic energy density is enhanced in this region due to the Kelvin-Helmholtz instabilities along the jet channel. 

The jet emission is primarily contributed by freshly-shocked particles with power-law spectra. A large contribution from these particles to the overall emission in the radio lobe will push the spectral break to higher frequencies, affecting the results of our spectral fitting. For the purposes of studying the spectral ageing of FR-II-like radio lobes, we reduce the jet emission by a factor of $100$ for consistency with observed FR-II-type radio sources. We define the jet as all Lagrangian tracer particles with a fluid tracer $> 0.2$ and a fractional time since last shock acceleration $< 4\%$. 

In Paper I, we noted that for simulations RAG-B327 and RAC-B327, BRAiSE produces sources with an FR index around $1.5$, indicating that the radio morphology is not FR-II like. With the reduction of the jet emission to $1\%$ of its full value, which reconciles our jet luminosities with typically observed values, the time-averaged FR index is $1.6 - 1.7$, indicating that the sources in simulations RAG-B327 and RAC-B327 appear to be more FR-II-like when the jet luminosity is reduced. 

\subsection{Spatially resolved spectral indices}
\label{section:spectral-index}

\begin{figure*}
    \centering
    \includegraphics{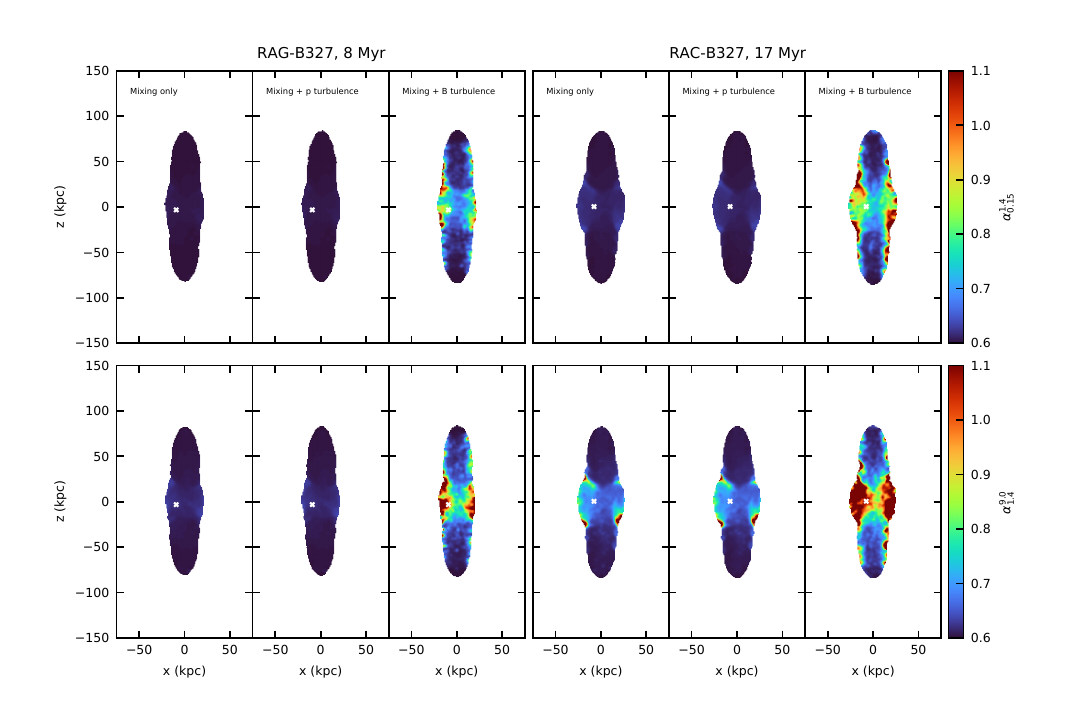}
    \caption[Spectral index maps for the three spectral models]{Spectral index maps for the spectral models for RAG-B327 (left) and RAC-B327 (right). The top row corresponds to the low frequency spectral index from $0.15 - 1.4$ GHz, and the bottom row corresponds to the high frequency spectral index from $1.4 - 9.0$ GHz. The spectral models from left to right are: mixing only, mixing with pressure turbulence, and mixing with magnetic turbulence. The jet emission has been downweighted by a factor of 100 in each case. The locations of the Tribble JP fits discussed in Section \ref{section:SED} are marked with white crosses.}
    \label{fig:alpha}
\end{figure*}

The spatially resolved spectral index between $0.15 - 1.4$ GHz and $1.4 - 9.0$ GHz is plotted for each of the emission models for simulations RAG-B327 and RAC-B327 in Fig. \ref{fig:alpha}. For both the group and cluster simulations, the `mixing only' and `mixing + pressure turbulence' models are broadly consistent with one another. In the group simulation, the spectral indices for these models are fairly flat across the source (i.e. close to the injected spectral index of $0.6$), with some steepening seen in the equatorial region for the high frequency spectral index, as expected for an FR-II radio lobe \citep{alexander_ageing_1987}. 

However, in the cluster simulation, steepening occurs in the equatorial regions for both the low and high frequency spectral indices. This is due to significant mixing between the shocked shell and cocoon in the dense cluster environment that inhibits the backflow and prevents younger particles from reaching the equatorial region (see Paper I for more details). Hence, the particles in the equatorial region are not mixing with younger electron populations as much and are steepening accordingly. Qualitatively, the spectral indices for the non-magnetic models have similar steepening to that shown in Figs. 10 and 11 (at the same redshift of $z = 0.05$) of \citet{yates-jones_praise_2022}, whereas this is not seen for the group environment. The radio sources shown in \citet{yates-jones_praise_2022} are propagating into a cluster environment, indicating that the different jet dynamics in the higher density environment influences the spectral steepening.

For both the group and cluster simulations, the `mixing + magnetic turbulence' model demonstrates significant spectral steepening in the equatorial regions and along the lobe edges for both the low and high frequency spectral indices, with spectral curvature developing between the low and high spectral index. The spatial distribution of spectral indices exhibits more small-scale structures than for the non-magnetic emission models due to the non-uniform distribution of magnetic fields in the lobe. This increases both the low and high spectral indices in the jet head region to be slightly steeper than the injection index. This steepening in the jet head region is greatest in the cluster environment, particularly for the high frequency spectral index.

The edges of the lobe have high magnetic energy density due to mixing between the lobe and the shocked shell that sweeps up and amplifies the ambient magnetic field \citep[Paper I,][]{jerrim_faraday_2024}. The volume filling factor of radiating particles is low in this mixing region (around $10^{-4}$), reducing the emission in this region but increasing the magnetic field strength of the radiating particles. This increase of high magnetic energy density at the lobe edges is seen in other jet simulations and is caused by shear at the contact surface between the jet cocoon and the ambient medium \citep{matthews_models_1990, huarte-espinosa_3d_2011}. However, we note that there is likely some level of numerical error in our sub-grid method. These lobe-edge particles with high magnetic energy densities are subject to greater synchrotron losses, thus steepening the spectral indices along the lobe edges towards the jet head.

The cluster simulation demonstrates greater steepening in the equatorial regions of the lobe than the group simulation for the `mixing + magnetic turbulence' model, which is likely due to the large population of older particle populations in this region as discussed above. An additional effect for this emission model is that the average magnetic field strength is higher for the lobes in the cluster environment (Fig. \ref{fig:lobeB-avg}), which therefore increases the spectral steepening relative to the lobes in the group environment.

\subsection{Spectral fitting}
\label{section:SED}

\begin{figure*}
    \centering
    \includegraphics{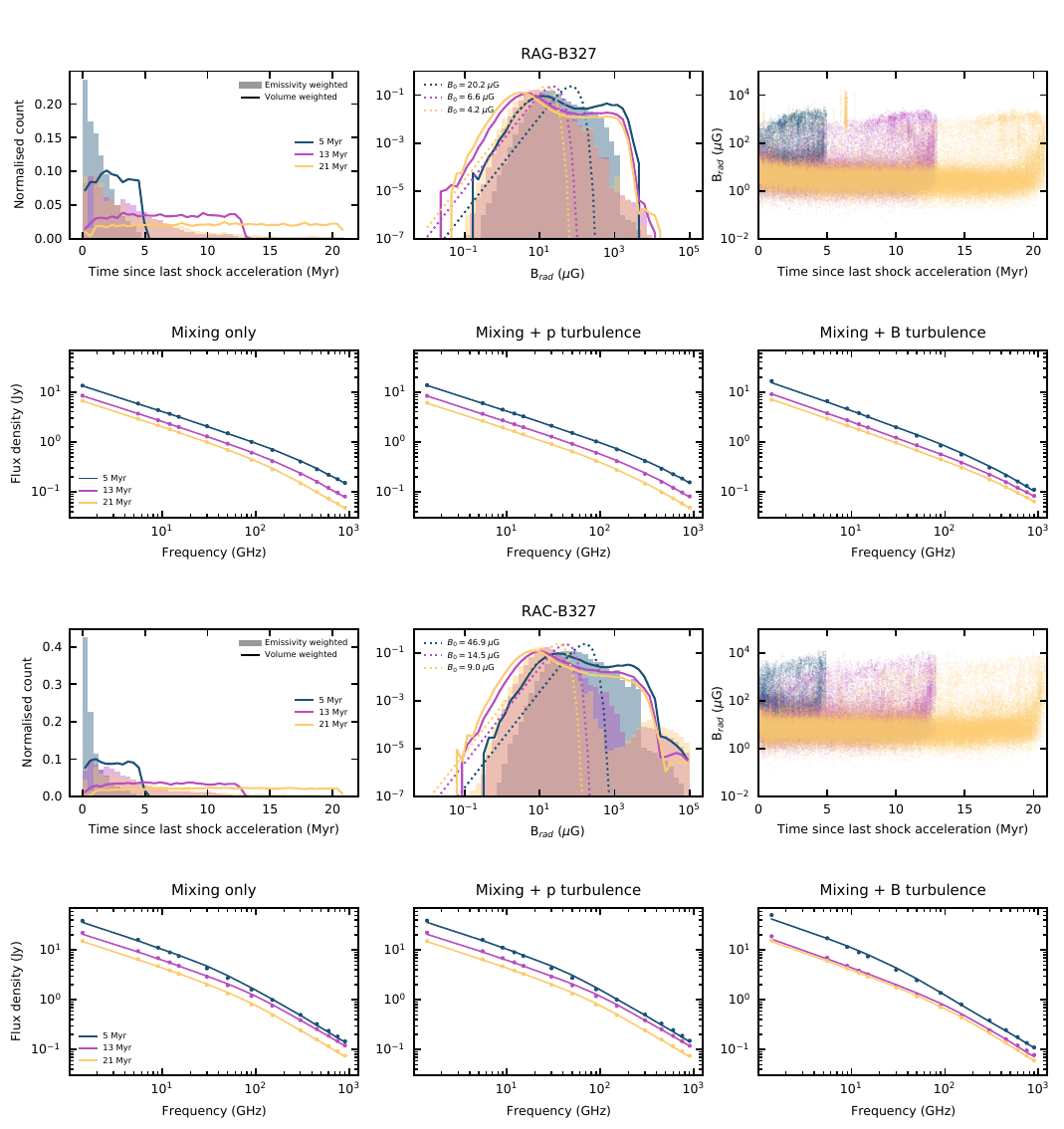}
    \caption[Particle ages, magnetic field strengths, and radio spectral energy distributions for the Tribble CI model]{Particle properties and radio spectral energy distributions for simulations RAG-B327 (top two rows) and RAC-B327 (bottom two rows) at $5$, $13$, and $21$ Myr. Rows 1 and 3: distributions of particle age (i.e., time since particle was last shocked), magnetic field strength, and the interaction between the two properties are shown from left to right. We plot both emissivity weighted (filled histogram) and volume weighted (solid line) distributions. For the magnetic field strength distribution panels, we include Maxwell-Boltzmann distributions (dotted line) with mean values equal to the lobe-averaged B field (Fig. \ref{fig:lobeB-avg}) at each time. Rows 2 and 4: Tribble CI fits computed using the \textsc{synchrofit} Python library \citep{quici_selecting_2022} for the three spectral models, from left to right. All spectra are fit at a redshift of $z = 0.05$.}
    \label{fig:CImodel}
\end{figure*}

\begin{figure*}
    \centering
    \includegraphics{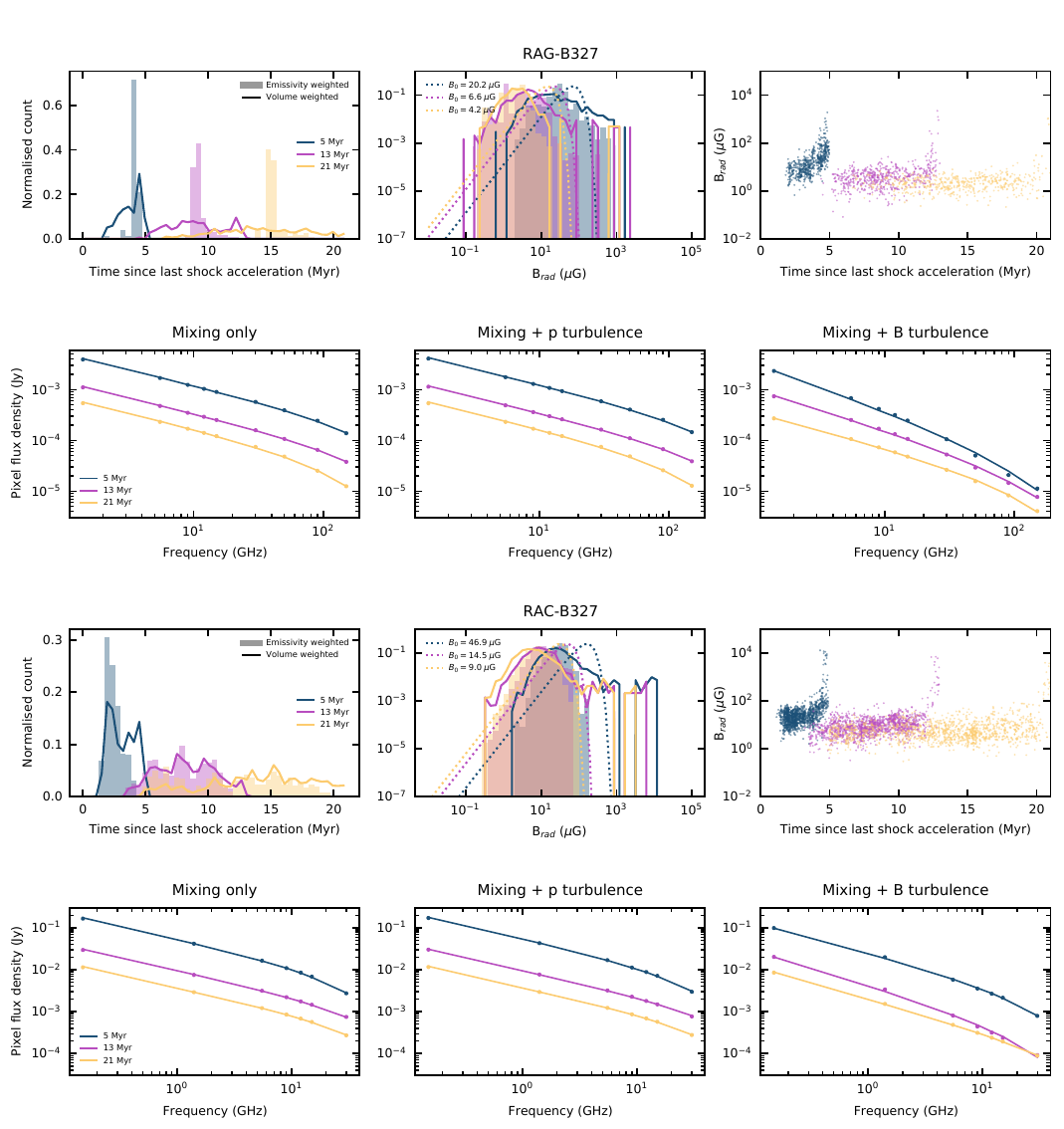}
    \caption[Particle ages, magnetic field strengths, and radio spectral energy distributions for the Tribble JP model]{Particle properties and radio spectral energy distributions as in Fig. \ref{fig:CImodel}, but corresponding to a single location in the equatorial region of the radio lobe for simulations RAG-B327 ($x = -9.2$ kpc, $z = -3.2$ kpc) and RAC-B327 ($x = -7.6$ kpc, $z = 0.4$ kpc). Instead of the Tribble CI model, the spectra are fit with Tribble JP models using the \textsc{synchrofit} Python library \citep{quici_selecting_2022} for the three spectral models, from left to right. All spectra are fit at a redshift of $z = 0.05$.}
    \label{fig:JPmodel}
\end{figure*}

To characterise the spectral energy distributions (SEDs) for each of our emission models, we use the \textsc{synchrofit} Python library \citep{quici_selecting_2022} to fit Tribble CI (TCI) models to the integrated lobe emission and Tribble JP (TJP) models to the lobe emission at specific resolved locations in Figs. \ref{fig:CImodel} and \ref{fig:JPmodel} respectively. We fit the break frequency in a range between $10^9$ and $10^{12}$ Hz at a redshift of $z = 0.05$. For the TCI models, we set the injection index to $0.6$ for the non-magnetic models. For the `mixing + magnetic turbulence' model, we allow the injection index to vary between $0.6 - 0.65$ to account for the steepening of the low-frequency spectral index. For both simulations, we fit the TCI models between $1.4 - 900$ GHz to accurately capture the break frequency.

\begin{figure*}
    \centering
    \includegraphics{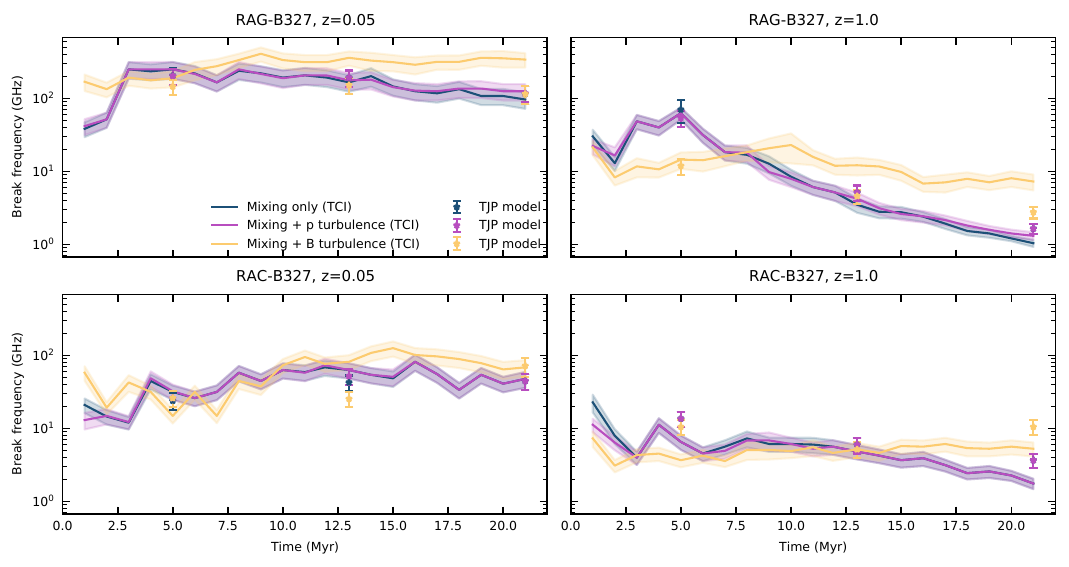}
    \caption[Break frequency over time]{Break frequency over time from the Tribble CI spectral fits for each of the emission models for simulations RAG-B327 (top row) and RAC-B327 (bottom row). Individual points shown as stars indicate the Tribble JP models fitted in Fig. \ref{fig:JPmodel}. The shaded regions indicate the $1\sigma$ uncertainty. The left column shows the spectral fits at $z = 0.05$ and the right column shows the spectral fits at $z = 1$.}
    \label{fig:breakfreq}
\end{figure*}

We calculate the TJP models at individual locations in each simulation (as shown in Fig. \ref{fig:alpha}) and set the injection index for the model to the lowest-frequency spectral index to improve the spectral fitting. In simulation RAG-B327, this is the $1.4 - 5.5$ GHz spectral index; for simulation RAC-B327, this is the $0.15 - 1.4$ GHz spectral index. For the group and cluster simulations respectively, we fit the TJP models between $1.4 - 150$ GHz and $0.15 - 30$ GHz to accurately capture the break frequencies in each case, since the break frequencies are higher in the group environment (Fig. \ref{fig:breakfreq}). We choose the individual locations in the equatorial region, where the oldest lobe emission is, as the TJP model assumes that the particles are injected at the same time. For simulation RAG-B327 this is at $x = -9.2$ kpc, $z = -3.2$ kpc, and for simulation RAC-B327 this is at $x = -7.6$ kpc, $z = 0.4$ kpc. The radio spectrum is produced by the particles along the line of sight within a beam FWHM ($3$ kpc) of these coordinates; at $5$/$13$/$21$ Myr, the number of particles that contribute to the spectrum is $549$/$547$/$507$ and $1228$/$1304$/$1109$ particles for the group and cluster simulations respectively. At these times, the total number of particles injected onto the simulation grid is $40186$/$104230$/$168276$ particles for both simulations.

In Paper I, the `mixing + magnetic turbulence' model was shown to have a $0.15$ GHz luminosity that was greater than the `mixing + pressure turbulence' model: $30\%$ and $40\%$ greater for the group simulation at $8$ Myr and the cluster simulation at $17$ Myr, respectively. This difference is shown in Fig. \ref{fig:CImodel}, which demonstrates that the `mixing + magnetic turbulence' models produce brighter radio sources than the non-magnetic models. For both simulations, there are minor differences in the shape of the SED for the `mixing only' and `mixing + pressure turbulence' models at each time shown. However, the `mixing + magnetic turbulence' model demonstrates that the turbulent magnetic field has a greater influence on the spectrum than particle mixing: the magnitude of the spectrum does not decrease over time as much as for the non-magnetic emission models. The distribution of magnetic field strengths (middle panels of rows 1 and 3 in Fig. \ref{fig:CImodel}) allows for higher magnetic field values than the non-magnetic models, which assume the average magnetic field strength applies across the entire lobe. The higher magnetic field strengths increase the overall luminosity in comparison to the non-magnetic models.

For both simulations, we find a general trend of increasing magnetic field strength for particles with times since they were last shock accelerated approaching the dynamical age of the source. This is shown in the panels in Fig. \ref{fig:CImodel} corresponding to the relationship between the magnetic field strength and the time since the particle was last shocked. These particles that have long times since they were last shock accelerated are generally distributed towards the edges of the lobes, which have low tracer values and therefore high magnetic field strength (as $B_{\rm rad} = B / \sqrt{\rm trc}$; see Equation \ref{eqn:volumefill_specage}). These particles make up the majority of the high tail of the volume-weighted magnetic field strength distribution ($B_{\rm rad} \gtrsim 10^3 \mu$G). As the emissivity-weighted magnetic field distribution shows, these particles do not contribute greatly towards the overall emissivity due to their low tracer values.

We demonstrate the effects of different particle age (i.e. time since the particle was last shocked) and magnetic field distributions on the spectrum of an individual pixel in Fig. \ref{fig:JPmodel}. Since there is a range of particle ages (for the volume-weighted distributions), the TJP model assumption of a single injection event does not hold. However, when weighted by the contribution to the overall emissivity of the radio lobe, we find that this assumption is not greatly violated for simulation RAG-B327. For both simulations, we once again see little difference in the SED and the TJP model fits for the `mixing only' and `mixing + pressure turbulence' emission models. However, we see a significant difference in the SED shape over time for the `mixing + magnetic turbulence' model once again due to the distribution of magnetic field strengths of the particles contributing to the emission at this location (i.e. within one beam FWHM of the central point). 

As seen in Fig. \ref{fig:CImodel} for the whole lobe, there is once again a general trend of increasing magnetic field strength with particle age in Fig. \ref{fig:JPmodel} for both simulations. Once again, this is due to the variations in fluid tracer value; along each line of sight, particles from each depth in the lobe will contribute. However, unlike for the whole lobe, the distribution of particle ages is more variable at the select locations for both simulations over time. We also note the lower amount of particles with strong magnetic fields $> 10^3 \mu$G for both simulations; the individual locations were chosen to not include any (reduced) jet emission, which contributes strongly to the high end of the magnetic field distribution in Fig. \ref{fig:CImodel}. Due to the lack of young particles with high magnetic field strengths, which contribute greatly to the spectrum (see Section \ref{section:discussion_specage}), the spectra in Fig. \ref{fig:JPmodel} show more steepening at lower frequencies than the integrated spectra over the whole lobe shown in Fig. \ref{fig:CImodel}.

The steepening of the spectrum is characterised by the spectral break frequency in the TCI and TJP models, which informs the estimated spectral age of the source. We plot the fitted break frequencies in Fig. \ref{fig:breakfreq} for each of our emission models, including the break frequencies of the fitted TJP models for the locations indicated in Fig. \ref{fig:JPmodel}. The TJP models are broadly consistent with the TCI model for both simulations. This indicates that at these locations of old emission, the magnetic field structure and particle mixing produces a spectrum with steepening (detected by the TJP model) that is representative of the overall source (as found by the TCI model).

Fig. \ref{fig:breakfreq} demonstrates that the break frequency for each of our models is always $> 10$ GHz; these break frequencies are higher than typically observed \citep[e.g.][]{turner_raise_2018_III}. The properties of our simulated lobes are such that the magnetic field strength of the radiating particles is generally close to $B_{ic}$ ($= 3.5 \mu$G at $z = 0.05$; compare to middle panels of rows 1 and 3 in Fig. \ref{fig:CImodel}). In this regime, the spectral break frequency approaches its maximum value at $\langle B \rangle \sim B_{ic}/\sqrt{3}$, assuming $B \sim \langle B \rangle$; the exact value will change slightly for a time-varying magnetic field. This is a coincidence of our chosen simulation parameters; if our simulations produced either a higher or lower average lobe magnetic field strength, the spectral breaks for our simulated sources would lie in the typically observed range ($\lesssim 10$ GHz). As the $B_{ic}$ value is redshift dependent, we study our simulated sources at a higher redshift of $z = 1$ so that the break frequency is reduced from its maximum value to that of typical observing frequencies.

The break frequency decreases significantly with redshift for our simulated sources: since $B_{ic} = 12.7 \mu$G at a redshift of $z = 1$, the magnetic field of the radiating particles is generally lower than $B_{ic}$, so the break frequency will be lower (see Equation \ref{eqn:spectral-age}). For both the TCI and TJP models at $z = 1$, we fit the spectra between frequencies of $0.15 - 9$ GHz as frequencies $> 9$ GHz experience significant radiative losses. The break frequencies for $z = 1$ are shown in Fig. \ref{fig:breakfreq}; in general, these frequencies are at least an order of magnitude lower than the estimated break frequencies at $z = 0.05$. The break frequencies for both simulations at $5$ Myr are still high as the magnetic field strength of the particles contributing to the emission are close to $B_{ic}$ (Fig. \ref{fig:JPmodel}). 

At high redshift, we find a significant decrease in the break frequency over time for the non-magnetic models in the group environment. This is due to the magnetic field strengths of the lobes in each environment: in simulation RAG-B327 the average lobe magnetic field strengths fall below $B_{ic} = 12.7 \mu$G within the first $10$ Myr, whereas the lobes in simulation RAC-B327 remain closer to this value over time (Fig. \ref{fig:lobeB-avg}). The magnetic models are impacted by the distribution of magnetic field strengths in the lobe, which allows for more particles to have a magnetic field strength close to the $B_{ic}$ value than in the non-magnetic models. In the group environment, the break frequency remains fairly consistent over time for the `mixing + magnetic turbulence' model. Therefore, the relative strengths of the lobe magnetic field and the magnetic field equivalent to the cosmic microwave background may have a strong impact on the observed break frequency.

\begin{figure*}
    \centering
    \includegraphics{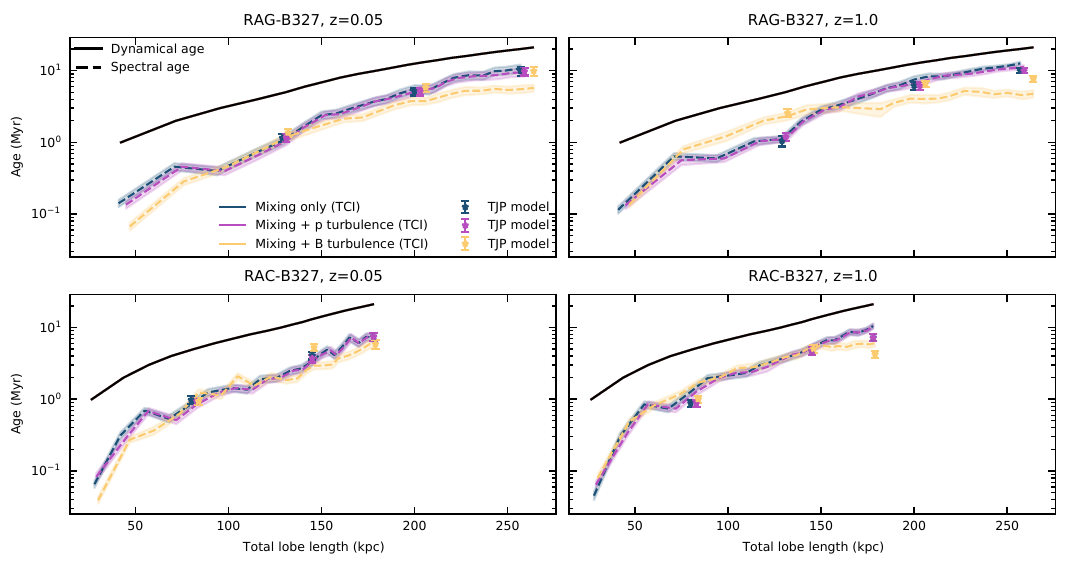}
    \caption[Spectral age vs lobe length]{Spectral ages and dynamical age vs length for RAG-B327 (top row) and RAC-B327 (bottom row). Spectral ages over time are calculated assuming the Tribble CI model. Individual points shown as stars indicate the Tribble JP models fitted in Fig. \ref{fig:JPmodel}. The shaded regions on the TCI models indicate the $1\sigma$ uncertainty. The left column shows the spectral ages at $z = 0.05$ and the right column shows the spectral ages at $z = 1$.}
    \label{fig:specvsdyn}
\end{figure*}

\subsection{Spectral ages}
\label{section:dyn-spec-ages}

For each of our spectral models, we derive spectral ages from the fits to the spectral energy distributions at low and high redshifts of $z = 0.05$ and $z=1$ respectively. The spectral ages plotted in Fig. \ref{fig:specvsdyn} are calculated from the break frequency from the TCI model fits. We use the volume-weighted average of the logarithm of $B_{\rm rad}$ as shown in Fig. \ref{fig:lobeB-avg} for the magnetic field strength for both models. The individual points in Fig. \ref{fig:specvsdyn} correspond to the TJP model fits to the spectra at $5$, $13$, and $21$ Myr for a single pixel.

The estimated spectral age for both the TCI and TJP models increases with the size of the radio lobe and approaches the simulation (dynamical) age as the lobe expands. The discrepancy between the spectral age and dynamical age is greater in the cluster environment, with spectral ages being a factor of $3.1$ lower at $21$ Myr at low redshift, compared to the group simulation, which is a factor of $2.4$ lower. The spectral ages at a higher redshift of $z = 1$ are closer to the dynamical age, however, there are some differences between the emission models. \citet{turner_raise_2018_II} suggest that the age discrepancy for the JP model will arise due to the mixing of particle ages, however, the CI model should reproduce the source dynamical age if the magnetic field strength is homogeneous. The TCI model assumes that there is a Maxwell-Boltzmann distribution of magnetic field strengths, however, as our lobe magnetic field strengths do not follow a Maxwell-Boltzmann distribution (Fig. \ref{fig:CImodel}), we do not reproduce the source dynamical age even with this model.

Given that the lobe magnetic field strengths do not follow a Maxwell-Boltzmann distribution, we estimate the magnetic field that should be assumed in the spectral model to obtain the true age. We find that the spectral age is more consistent with the true age if the average magnetic field strength as shown in Fig. \ref{fig:lobeB-avg} is reduced by a factor of $2.09$ or $4.46$ for simulations RAG-B327 and RAC-B327 respectively. These reduced magnetic field estimates approximately correspond to the 25th percentile and 7th percentile of the total lobe magnetic field distribution for simulations RAG-B327 and RAC-B327 respectively, indicating that the lower end of the magnetic field distribution produces a spectral age estimate consistent with the true age. The lower end of the emissivity-weighted magnetic field distribution (i.e. the `visible' magnetic field) is generally underpredicted by the Maxwell-Boltzmann distribution, which could pose a significant problem for observers to confidently fit. Further work is needed to compare the observed magnetic field strengths \citep[e.g. X-ray inverse-Compton measurements;][]{ineson_representative_2017} to the point in the magnetic field distribution that is relevant for ageing models.

The age calculated at $21$ Myr by the individual fits to the TJP models at $z = 0.05$ for the `mixing only' and `mixing + pressure turbulence' models is $10 \pm 1$ Myr for simulation RAG-B327 and $7.5 \pm 0.9$ Myr for simulation RAC-B327. By using the magnetic field to calculate the radio emission, this age estimate stays at $10 \pm 1$ Myr for RAG-B327 and decreases to $5.9 \pm 0.9$ Myr for RAC-B327. In the cluster environment, the magnetic field strength is higher (Fig. \ref{fig:lobeB-avg}), so the lobe emission is much dimmer at high frequencies and the oldest lobe plasma will be contributing less to the overall spectrum. Therefore, the younger particles will dominate the spectrum at this location, reducing the spectral steepening and decreasing the spectral age estimate.

At a redshift of $z=1$ in the group environment, the spectral ages derived from TJP model fits for the non-magnetic models are broadly consistent with those from the TCI model fits, but for the magnetic emission model they are up to a factor of $1.6$ higher. In the cluster environment, the spectral ages derived from the TJP model fits are up to a factor of $1.7$ lower than the TCI equivalent, which occurs for the `mixing + magnetic turbulence' model at $5$ Myr. These differences are due to the distribution of magnetic field strengths at the location of the TJP model fits; e.g. for the cluster environment, Fig. \ref{fig:JPmodel} shows that the magnetic field strength of particles at this location is peaked close to the $B_{ic}$ value at $z = 1$, indicating that the spectral break frequency is close to its maximum value, and therefore the spectral ages are close to their minimum value at this location. However, for the group environment at both low and high redshift, the TJP spectral ages are closer to the true dynamical age as the spectrum at this location has a greater contribution from older particles than the overall spectrum that the TCI model is fit to.

The behaviour of the spectral age estimate for the `mixing + magnetic turbulence' model as the source grows in size is significantly different to the non-magnetic emission models at this higher redshift. For both the group and cluster environments, the spectral age for the non-magnetic emission models continues to increase as the break frequency decreases. However, the `mixing + magnetic turbulence' model does not have a significant decrease in the spectral break frequency due to the inhomogeneous magnetic field: the overall spectrum is flatter due to the range of particle magnetic field strengths contributing to it (see further discussion in Section \ref{section:discussion_specage}). The magnetic field therefore reduces the spectral age estimate at a redshift of $z = 1$ for our simulated sources.

It is clear that each of the spectral models explored in this paper do not provide an accurate estimate of the dynamical age of the lobes in both our group and cluster simulations. The difference between the spectral age and the true dynamical age varies with lobe length, redshift, and spectral model, however, in the cases studied here, the ages of the lobes are underestimated by at least a factor of $2$. Therefore, if the power output of a particular radio-loud AGN is known and its spectral age has been estimated, then the estimated energy deposited into the environment of the AGN could be half (or less) than the true amount of energy that would be available for feedback. This has significant implications for our understanding of galaxy formation and evolution, and the role that AGN feedback plays in these processes.

\section{Discussion}
\label{section:discussion_specage}

\begin{figure*}
    \centering
    \includegraphics{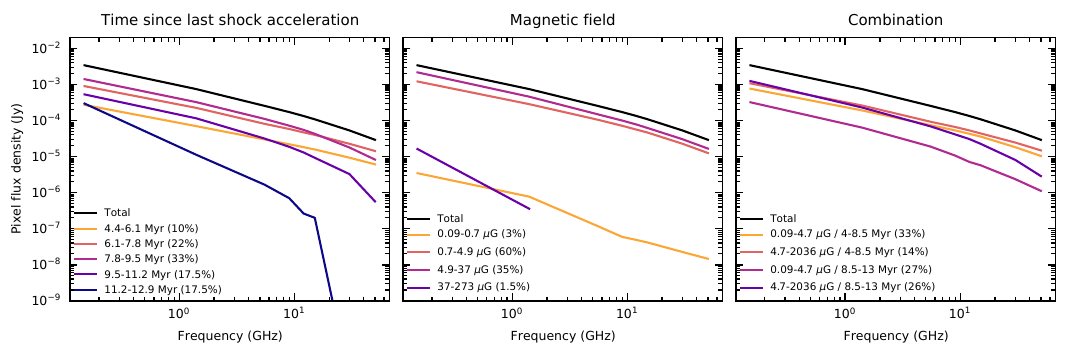}
    \caption[Contributions to radio spectrum of a single pixel]{Spectrum of a single pixel ($x = -9.2$ kpc, $z = -3.2$ kpc) for simulation RAG-B327 at $13$ Myr, with different age and magnetic field strength bins to demonstrate the contribution of each population. Left panel: binning over particle ages (i.e. time since the particle was last shock accelerated). Centre panel: binning over particle magnetic field strengths (note: from $273 - 2036 \mu$G, there is no emission). Right panel: binning over two age/magnetic field strength categories. The percentages refer to the number of particles in each bin.}
    \label{fig:binspec}
\end{figure*}

The distribution of both the turbulent lobe magnetic field and the age of the radiating particles will affect the synchrotron radio spectrum resulting from these particles. In Fig. \ref{fig:binspec}, we plot the spectrum of a single pixel from simulation RAG-B327 at $13$ Myr, and the contributions to this spectrum from different particle ages and magnetic field strengths. 

The left panel of Fig. \ref{fig:binspec} demonstrates how the youngest electrons at this location dominate the high-frequency curvature of the spectrum: despite the majority of particles at this location having times since they were last shock accelerated between $7.8-9.5$ Myr, the overall spectrum does not show the significant steepening that this population has. This steepening is offset by the younger particles. The oldest particles show a significant reduction in flux density at high frequencies, but this is not reflected by the overall spectrum. Therefore, the SED is most representative of the younger populations of particles, and the TJP model will underpredict the spectral age estimate \citep[as shown by][]{turner_raise_2018_III}. 
\\

In the centre panel of Fig. \ref{fig:binspec}, we see that $95$ percent of the emission is provided by particles with magnetic field strengths between $0.7 - 37 \mu$G. We note that at this location, $0.5$ percent of the particles have magnetic field strengths of $273 - 2036 \mu$G, however, they are not emitting as they are all $> 12$ Myr old and hence are subject to severe radiative losses. The particles with magnetic field strengths $4.9 - 37 \mu$G demonstrate a steeper spectrum than the total spectrum, indicating that the distribution of magnetic field strengths will impact the interpretation of the spectral age.

The right panel of Fig. \ref{fig:binspec} shows the interaction of the effects of particle age and magnetic field strength. For both low and high magnetic fields, we see that the older population has significant steepening in the spectrum that is not reflected in the overall shape of the spectrum. At all frequencies, the greatest contribution to the overall flux density is from the higher magnetic field strength population; at high frequencies, it is the young particles that are contributing the most emission. This is despite the fact that this population of young particles with high magnetic field strength is only $14$ percent of the total population of particles contributing to the spectrum at this location. 

The spectrum at a given location in the radio lobe will be dominated by its youngest particles. The effect of age is more significant than the effect of magnetic field in an individual sense, however the youngest particles with higher magnetic field strengths contribute a significant amount of emission to the overall spectrum, indicating that the relationship between particle age and radio SED is affected by the distribution of magnetic field strengths in the emitting region.

\section{Conclusions}
\label{section:conclusions_specage}

In this paper we have used three-dimensional relativistic magnetohydrodynamic simulations of FR-II-like radio galaxies to study the effects of particle mixing and turbulent magnetic fields on the estimated spectral age of radio galaxies. We compare the estimated spectral ages for three emission models: mixing of radiating particles only, mixing with non-uniform pressures, and mixing with turbulent magnetic fields. We demonstrate that using the turbulent magnetic field present in the radio lobe to calculate the synchrotron emission does not greatly impact the spectral age estimate, suggesting that mixing of particle populations with different ages is the primary source of the discrepancy between spectral age estimates and the true dynamical age. At a higher redshift ($z=1$), we find that the turbulent magnetic field reduces the spectral age estimate with respect to the non-magnetic emission models, however particle mixing is still the dominant effect in this case.

We find that for our three emission models, using the turbulent magnetic fields makes a significant difference to the surface brightness morphology, spectral indices, and spectral energy distribution in both group and cluster environments. We find that the turbulent magnetic fields present in our radio lobes increase the steepening of the radio spectrum of the lobes, compared to our non-magnetic models. This effect is demonstrated by spatial spectral indices for both low and high frequency: the turbulent magnetic fields produce a clumpy structure throughout the lobe, and increase the steepening of the spectrum towards the equatorial regions of the lobes. 

We use the \textsc{synchrofit} Python library \citep{quici_selecting_2022} to fit Tribble continuous injection (TCI) models to the spectra of our simulated FR-II-like radio lobes and Tribble Jaffe-Perola (TJP) models to individual locations in the oldest parts of our lobes. We find that the break frequencies fitted to our spectra at a redshift of $z = 0.05$ are higher than $10$ GHz, because the lobe magnetic fields are comparable to the magnetic field strength equivalent of the cosmic microwave background. At a higher redshift of $z = 1$, the break frequencies are significantly reduced, since the magnetic field strength equivalent of the cosmic microwave background is higher than the majority of the radiating particles that contribute to the emission.

We examine the interacting effects of particle age and magnetic field strength on the spectrum at a single location in our group simulation at low redshift. The shape of the spectrum is primarily determined by young particles with high magnetic field strengths, which comprise a small percentage of the overall particles at a specific location. The young particles set the curvature of the spectrum, whereas the particles with high magnetic field strengths set the normalisation of the spectrum. 

The spectral ages estimated for each of our emission models over time are consistently lower than the true dynamical age. We show that the TJP model assumption of a single particle age at a particular location is not met due to particle mixing, and that the TCI model assumption of a Maxwell-Boltzmann distribution of magnetic field strengths is in general not met due to dynamical processes affecting the magnetic field. Therefore, neither of these models can provide an accurate estimate of the dynamical ages of our simulated sources, which has significant implications for estimates of AGN energetics. Knowledge of the turbulent dynamics in AGN jet-inflated cocoons is therefore needed to accurately interpret the spectra of radio galaxies. In future work, we will explore removing the assumptions of these models and thereby improving the spectral age accuracy.

\begin{acknowledgement}

We thank the anonymous referee for a high-quality, constructive report which has improved this paper. LJ thanks the University of Tasmania for an Australian Government Research Training Program (RTP) Scholarship. SS, CP and PYJ acknowledge funding by the Australian Research Council via grant DP240102970. This research was carried out using the high-performance computing clusters provided by Digital Research Services, IT Services at the University of Tasmania. This work has made use of data from The Three Hundred collaboration (https://www.the300-project.org) which benefits from financial support of the European Union’s Horizon 2020 Research and Innovation programme under the Marie Skłodowskaw-Curie grant agreement number 734374, that is the LACEGAL project. The Three Hundred simulations used in this paper have been performed in the MareNostrum Supercomputer at the Barcelona Supercomputing Center, thanks to CPU time granted by the Red Española de Supercomputación. We thank \citet{knapen_how_2022} for their guide to writing astronomy papers. We acknowledge the support of the developers providing the Python packages used in this paper: Astropy \citep{astropy_collaboration_astropy_2022}, JupyterLab \citep{kluyver_jupyter_2016}, Matplotlib \citep{hunter_matplotlib_2007}, NumPy \citep{harris_array_2020}, and SciPy \citep{virtanen_scipy_2020}.

\end{acknowledgement}





\paragraph{Data Availability Statement}

The data underlying this article will be shared on reasonable request to the corresponding author.


\bibliography{references}

\appendix

\newpage

\section{Comparison of JP, TJP, CI, and TCI models to broken power-law fits}
\label{section:appendix}

\begin{figure}
    \centering
    \includegraphics{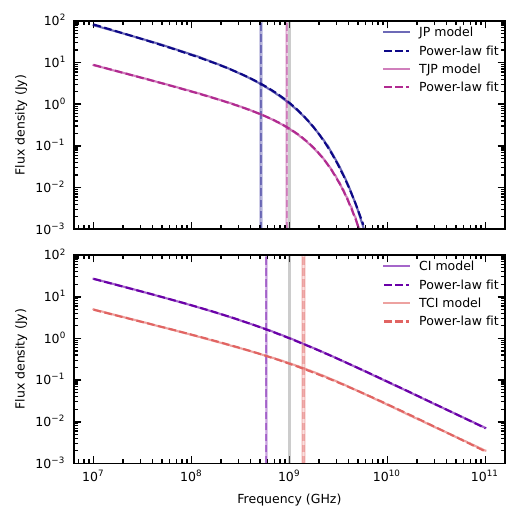}
    \caption{JP/TJP (top) and CI/TCI (bottom) spectra from \textsc{synchrofit} fit with the broken power-laws from Equations \ref{eqn:jppowerlaw} and \ref{eqn:cipowerlaw}. The vertical solid grey line corresponds to the break frequency of the generated spectra ($1$ GHz) and the dashed lines correspond to the estimated break frequency for the broken power-law fits. The shaded region around the dashed lines correspond to the first standard deviation of these values. Each model has been normalised such that the flux density at $1$ GHz is $1$ Jy and the sensitivity is $1$ mJy. The Tribble models have been decreased by a factor of $4$ so that each spectrum is clearly visible on the plot.}
    \label{fig:brokenpowerlaw}
\end{figure}

In Fig. \ref{fig:brokenpowerlaw} we have used \textsc{synchrofit} \citep{quici_selecting_2022} to generate generic JP, TJP, CI, and TCI spectra and fit them with broken power-law methods to demonstrate the differences between these approaches for AGN radio lobe spectra fitting. For the CI and TCI spectra, we fit the following equation \citep{MrMoose}:

\begin{equation}
\label{eqn:cipowerlaw}
    {\rm log}_{10}(S_\nu) = {\rm log}_{10}(N) + a_1{\rm log}_{10}(\nu/\nu_b) + (a_2-a_1){\rm log}_{10}(\nu/\nu_b),
\end{equation}

where $N$ is the normalisation, $\nu_b$ is the spectral break frequency, $a_1$ is the low-frequency power-law, and $a_2$ is the high-frequency power-law. For the JP and TJP spectra, we fit the following equation:

\begin{equation}
\label{eqn:jppowerlaw}
    {\rm log}_{10}(S_\nu) = {\rm log}_{10}(N) + a_1{\rm log}_{10}(\nu/\nu_b) - {\rm log}_{10}\bigg(1 + e^{(\nu/\nu_b)^{|a_2 - a_1|}}\bigg).
\end{equation}

As shown in Fig. \ref{fig:brokenpowerlaw}, the estimated break frequency from these power-law methods is in general significantly different than the true break frequency ($1$ GHz), which translates to the estimated spectral ages for each model. The spectral age of each of the generic spectra is $4.04$ Myr; the JP and CI models estimate a spectral age of $5.6$ and $5.3$ Myr respectively (a fractional difference of $30\%$ or $40\%$), whereas the TJP and TCI models estimate spectral ages of $4.2$ and $3.4$ Myr respectively (a fractional difference of $3\%$ or $-15 \%$). Therefore, the relationship in Equation \ref{eqn:spectral-age} cannot be used with a break frequency derived from a broken power-law model to estimate the spectral age of a source accurately. 

Additionally, as discussed in Sections \ref{section:dyn-spec-ages} and \ref{section:discussion_specage}, the estimated spectral ages from (T)JP and (T)CI models in general cannot accurately estimate the true age of an AGN radio lobe as the model assumptions are not met (there is no single particle age for the JP model due to particle mixing, the CI model does not account for adiabatic losses, and the magnetic field does not follow a Maxwell-Boltzmann distribution as assumed by the Tribble models), thereby compounding the problem. We do not recommend that a spectral age of an AGN radio lobe be estimated using a broken power-law fitting method.

\end{document}